\documentclass[showpacs,amsmath,amssymb,twocolumn,floatfix,prl]{revtex4}
\usepackage[dvips]{graphicx}
\input{epsf}
 
\def\cnot{\textrm{C\scshape not}}

\begin{document}

\title{Intermediate quantum maps for
quantum computation} 

\author{O. Giraud and B. Georgeot} 
\affiliation{Laboratoire de Physique Th\'eorique, UMR 5152 du CNRS, 
Universit\'e Paul Sabatier, 31062 Toulouse Cedex 4, France}

\date{April 29, 2005} 

\begin{abstract}
We study quantum maps displaying spectral
statistics intermediate between Poisson and Wigner-Dyson.  It is shown
that they can be simulated on a quantum computer
with a small number of gates, 
and efficiently yield information 
about fidelity decay or spectral statistics.
We study their matrix elements
and entanglement production, and show that they converge
with time to distributions which differ from random matrix predictions.
A randomized version of these maps can be implemented
even more economically, and yields pseudorandom operators
with original properties, enabling for example to produce
fractal random vectors.  These algorithms
are within reach 
of present-day quantum computers.

\end{abstract}
\pacs{03.67.Lx, 03.67.Mn, 05.45.Mt}
\maketitle

The study of quantum information has attracted
more and more interest from the scientific community in the recent past.
Quantum communication and
quantum computation have been shown to be deeply different from their
classical counterparts.  Algorithms have been built 
showing that quantum computers can outperform classical devices for some 
problems \cite{Nielsen}.  In particular,
quantum mechanical systems can be simulated much faster on a quantum computer
\cite{lloyd}. An especially congenial class of models corresponds to
quantum maps.  They can display complex dynamics, but can be
described by simple evolution operators.  It has been shown that the
quantum baker's map \cite{schack}, the quantum kicked rotator \cite{GS},
the quantum sawtooth map \cite{complex} can be simulated efficiently
on a quantum computer.  Still, even the simplest models are difficult
to implement on the small-size quantum computers experimentally available,
and only the quantum baker's map has been implemented to date
with three qubits \cite{cory}.  General algorithms 
have been proposed to probe phase space distributions
\cite{wigner}, fidelity decay \cite{fidelity}, 
form factors \cite{formfactor} or localization length \cite{loclength}
of such maps. Quantum maps have been also used
as testgrounds to study the production of entanglement in quantum
systems \cite{caves}.  Very recently, they have been used 
as models to build pseudorandom operators which can be
efficiently implemented on quantum computers \cite{emerson}.

The models envisioned so far correspond mostly to chaotic maps, where the most
complex behavior is expected to manifest itself, and 
eigenvalue
statistics are close to those of the Wigner-Dyson ensembles
of Random Matrix Theory (RMT).  Some integrable maps, which are
expected to follow Poissonian statistics, have been also studied.
However, it has been observed recently that some systems obey 
{\it intermediate statistics}, 
characterized by a level
repulsion and a Poisson-like behavior at long distance in energy spacings.  
They can be modeled by the so-called semi-Poisson 
statistics \cite{gerland} for which correlation functions
can be explicitly calculated, giving
energy level spacing distributions
$P_{\beta}(s)\propto s^{\beta}e^{-(\beta+1)s}$. 
Such distributions have been first observed
 at the Anderson metal-insulator transition
for electrons in disordered systems \cite{braun}, 
and later in pseudointegrable systems \cite{gerland}
or in certain diffractive billiards
\cite{diffractive}.  They are usually associated with fractal properties of
eigenstates.

In this paper, we study from the viewpoint of quantum computation
a one-parameter family of quantum maps recently
introduced \cite{giraud}.  They can be expressed
in a particularly simple way, yet 
 the spectral statistics display a wide range of semi-Poisson 
distributions depending on the value of the parameter.
We first show that such maps
can be simulated efficiently on a quantum computer, and can be 
implemented with a remarkably small number of qubits and quantum gates.
In particular, it represents an ideal testground for algorithms proposed
in recent years which aim at measuring the form factor or the fidelity decay
using only one qubit of quantum information.  The properties
of these maps being intermediate between chaos and integrability, we 
study how this translates in the matrix element distributions
and the entangling power.  At last, we show that
a certain generalization of these maps which corresponds to
a new ensemble of random matrices with semi-Poisson
statistics recently proposed in \cite{bogomolny}
can be implemented even more efficiently on a quantum computer. They 
can be used
as a way to produce a family of
pseudorandom operators with new properties related to fractal behavior. 
We note that different distributions 
interpolating between Poisson and Wigner-Dyson, built from
a partial randomization procedure,
were studied in \cite{hellberg}.  

\begin{figure}
\begin{center} 
\vspace{0.5cm}
\includegraphics[width=.65\linewidth, angle=-90]{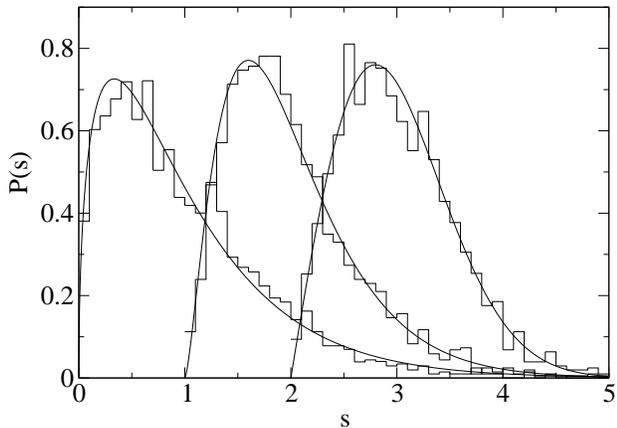} 
\end{center} 
\caption{Nearest-neighbor distribution of eigenvalues
of $\hat{U}$ for $N=2^{12}$ and from left to right
$\alpha=1/3, \alpha= 1/5, \alpha=(1+\sqrt{5})/2$. 
Data are taken from the dissymmetrized 
spectra of $\alpha$  and $1-\alpha$. 
Distribution corresponding to $\alpha=1/5$ (resp. $\alpha=(1+\sqrt{5})/2$)
is shifted by 1 (resp. 2) along the $s$-axis. 
The solid curves from left to right correspond to semi-Poisson with
$\beta=1/2$, $\beta=3/2$ and COE.}
\label{p(s)}
\end{figure}

We start with the classical map defined by 
$\bar{p} = p + \alpha \;\mbox{(mod} 
\;\mbox{1)}\;;\;\; 
\bar{q} = q+ 2\bar{p} \;\mbox{(mod} 
\;\mbox{1)}\;$
where $(p,q)$ is the pair of conjugated momentum (action) 
and angle 
variables, and the bars denote the resulting variables after one iteration 
of the map. 
The quantization of this map yields a unitary evolution operator
which can be expressed in momentum space by the $N \times N$ matrix
$U_{pp'}=\frac{\exp(-2i\pi p'^2/N)
\left(1-\exp(2i\pi N \alpha)\right)}{1-\exp (2i\pi (p-p'+N\alpha)/N)}$ 
\cite{giraud}, or alternatively in operator notation
$\hat{U} = e^{-2i\pi\hat{p}^2/N} e^{2i\pi \alpha \hat{q}}$.
For generic irrational $\alpha$, the spectral statistics of $\hat{U}$
are expected to follow RMT (in this case COE).
For rational $\alpha=a/b$, a variety of different behaviors are
observed \cite{giraud}.  In particular, it is conjectured from \cite{bogomolny}
that for $aN=\pm 1 \;\mbox{(mod} \;\mbox{b)}\;$   the statistics 
of eigenvalues is of the semi-Poisson type $P_{\beta}(s)$
with parameter $\beta =b/2 -1$.  
This can be checked in Fig.\ref{p(s)}
where the statistics of eigenvalues is plotted for different values
of $\alpha$ (for $N = 0 \;\mbox{(mod} \;\mbox{4)}\;$ there is an additional 
symmetry
$US=SU$, where $S_{qq'}=(-1)^q\delta_{qq'}$; each half of the spectrum
should be considered separately \cite{giraud}).
Thus $\hat{U}$ gives a set of quantum maps with statistics
corresponding to natural 
intermediate distributions between Poisson and RMT, in a controllable manner.

The map $\hat{U}$ can be implemented efficiently on a quantum computer.
Indeed, the use of the Quantum Fourier Transform (QFT) allows to 
switch from position to momentum representation.
More precisely, for one iteration of
$\hat{U}$ on a $N$-dimensional Hilbert space with
$N=2^{n_q}$, one first implements $\exp(2i\pi \alpha \hat{q})$,
which is diagonal in the $q$-representation.
This can be done efficiently
using the binary decomposition of $q$: if
$q=\sum_{j=0}^{n_q-1} q_j 2^j$, then
$\exp(2i\pi \alpha \hat{q})$ corresponds to the application
of the $n_q$ one-qubit gates $|0 \rangle \rightarrow |0\rangle$, $|1\rangle
\rightarrow 
\exp(2i\pi \alpha 2^j ) |1\rangle $. Then by using a QFT one can shift from
$q$ to $p$ representation, using 
$n_q(n_q+1)/2$ gates.
 In this representation, the second operator $ e^{-2i\pi\hat{p}^2/N}$
is diagonal. If $p=\sum_{j=0}^{n_q-1} p_j 2^j$, then
$\exp(-2i\pi p^2/N)=\prod_{j_1,j_2} \exp(-2i\pi p_{j_1} p_{j_2} 
2^{j_1+j_2}/N)$. To simulate it, one needs
$n_q^2$ two-qubit gates applied to each qubit pair 
$(j_1,j_2)$, keeping the states $|00\rangle, |01\rangle, |10\rangle$ 
unchanged while $|11\rangle \rightarrow \exp(-2i\pi 2^{j_1+j_2}/N) |11\rangle$.
Then a QFT brings back the wavefunction to
the $q$ representation.  In total, the evolution requires
$2n_q^2 +2n_q$ gates to be implemented, of which $2 n_q^2 - n_q$
are two-qubit gates. This is less than any other map 
proposed to date (including the sawtooth map \cite{complex}),
except the quantum baker's map,
which
has already been implemented \cite{cory}.

The algorithm above can be used as subroutine of other algorithms
which aim at measuring quantum-mechanical quantities, with the
attractive feature that it makes them very economical since the map
evolution needs remarkably little quantum gates.  One can for example 
probe Wigner and Husimi phase space distribution functions \cite{wigner}, 
or investigate fidelity decay in presence of perturbation 
\cite{fidelity}; the fact that these maps correspond to spectral statistics
intermediate between Wigner-Dyson and Poisson should translate into 
specific properties for these quantities.
It has also been proposed to use a quantum computer to differentiate between
quantum chaos and integrability by evaluating the form factor at short times
\cite{formfactor}.  In our case, the same algorithm can give much more
information.  The method in \cite{formfactor} adds
one probe qubit to the system, performs $U^n$,
and uses two additional one-qubit  gates
to transfer the trace of $U^n$ to the probe qubit.  Using such scattering 
circuits, real and imaginary parts of $\mbox{Tr}U^n/N $ correspond to
expectation values of Pauli operators for the probe qubit.  For COE or CUE, 
this quantity is of order $1/N$ for small $n$.  In the case of 
intermediate statistics ($\alpha=a/b$), 
one expects $<\mbox{Tr} U^n>_n \sim \kappa\sqrt{N}$, where the average is taken
over the first iterates of $U$. The form factor at short times 
is then given by $|\kappa|^2$. In order to get the value of $\kappa$
with enough precision,  one needs a number of quantum measurements 
of order $N$ (the number of values of $n$ to average over depends
only on $b$ \cite{giraud} and does not vary with $N$).
This implies a quadratic gain over classical computation.
For $N=0 \;\mbox{(mod} \;\mbox{4)}\;$, 
in order to dissymmetrize
the spectrum, one additionally has to perform the evolution of $SU^n$
(this only requires one extra controlled-phase gate), and
the difference
between the traces of $U^n$ and $SU^n$ gives the required quantity.
Using this algorithm for large $N$ enables to probe the form
factor at increasingly short times and check the
semiclassical conjectures \cite{giraud}.
The value of $|\kappa|^2$ also yields the level
compressibility, which controls the behavior of the spectral
number variance at infinity.
It is possible that an exponential
gain can be obtained through investigation of the fidelity decay
\cite{fidelity}.  We note that the fidelity decay can be explicitly
related to the form factor \cite{klaus}.
Both quantities can be probed using deterministic 
quantum computation with one single pseudopure bit \cite{laflamme},
which together with the small number of gates needed should make
 these simulations very attractive for NMR quantum computation.

\begin{figure}
\begin{center} 
\vspace{0.5cm}
\includegraphics[width=.65\linewidth, angle=-90]{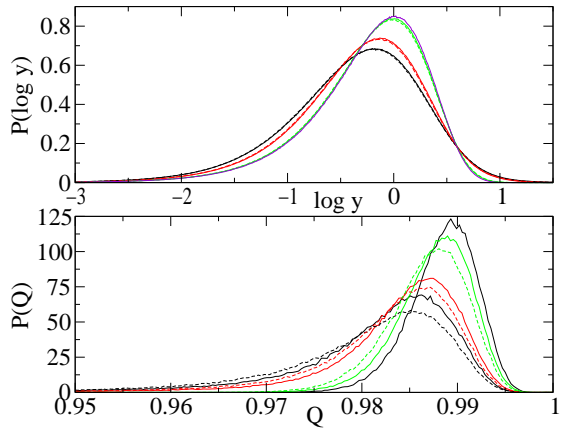} 
\end{center} 
\caption{(color online) Distributions
of matrix element (top) and $Q$ of column vectors
(bottom) for iterates $\hat{U}^n$ of $\hat{U}$ for $N=2^9$, in
$p$ representation.
Matrix elements $x$ are rescaled by $y=N|x|^2$.
Only one symmetry class of $S$ is taken. From lowest to topmost curve:
black is $\alpha=1/3$,
red is $\alpha=1/5$, green is $\alpha=(1+\sqrt{5})/2$, purple is CUE.
Full lines are averaged over  $10^5\leq n < 10^5 +1000$, 
dashed line are averaged over $1000 \leq n < 2000$ (top) and 
$2000 \leq n < 3000$ (bottom). Logarithms are decimal.}
\label{entangp2}
\end{figure} 

It is instructive to study the entangling power of these maps.  Indeed,
entanglement is a key resource for quantum information \cite{Nielsen}.
Quantum chaotic evolutions have been shown 
to generate entanglement distributions
similar to the predictions of RMT
\cite{caves}. The entanglement can be
measured by the average bipartite entanglement between one qubit and the 
rest of the system $Q=2-2/n_q\sum_{k=1}^{n_q} \mbox{Tr} \rho_k^2$, 
where $\rho_k$ is the density operator corresponding to the $k$-th 
qubit after having traced out the rest \cite{entanglement}.
Fig.\ref{entangp2} displays the distribution of matrix elements 
and $Q$ for the column vectors of $U^n$ for large $n$.
It is compared with the RMT prediction computed numerically
from the parametrization 
method of \cite{poles}.
For irrational $\alpha$, the spectral statistics follow COE,
but both matrix element and $Q$ distributions
converge to CUE predictions.
For values of $\alpha$ with intermediate statistics, there is a convergence
to a distribution different from CUE predictions.  This holds 
in $p$ representation: in  $q$ representation both quantities converge 
to CUE predictions (data not shown).
In all cases the
 convergence is faster for matrix elements than for $Q$; the convergence 
rate is
similar to what is obtained
 for chaotic maps \cite{caves}. 

In \cite{emerson}, pseudorandom operators
inspired by quantum chaotic maps were built which efficiently create 
entanglement and matrix element distributions close to RMT.  
In our case, it is possible to construct ensembles of Intermediate 
Statistics Random Matrices (ISRM)
based on the quantum map $\hat{U}$ \cite{bogomolny}. 
The procedure consists in replacing
$e^{-2i\pi\hat{p}^2/N}$ in $\hat{U}$ by the diagonal matrix 
(in $p$ representation)
$(e^{i\Phi_p}\delta_{pp'})$, where $\Phi_p$ are random variables
either independent
(``non-symmetric case'') or verifying
$\Phi_{N-p}=\Phi_p$ (``symmetric case'').  
General arguments \cite{bogomolny} indicate that for $\alpha$ irrational
eigenvalue statistics for ISRM follow COE (symmetric case) or CUE 
(non-symmetric case).  For rational $\alpha=a/b$, and matrix size
$N\rightarrow \infty$ obeying $aN=\pm 1 \;\mbox{(mod} \;\mbox{b)}\;$,
eigenvalue statistics follow the semi-Poisson prediction with
$\beta=b-1$ (non-symmetric case) or $\beta=b/2-1$ (symmetric case).
These ISRM can be implemented on a 
quantum computer, actually more economically than $\hat{U}$. 
The only difference with the algorithm simulating $\hat{U}$ consists in the
replacement of $e^{-2i\pi\hat{p}^2/N}$
by $(e^{i\Phi_p}\delta_{pp'})$.
This operator 
multiplies each basis state $|p\rangle$ by a Gaussian random phase. 
It can be simulated by choosing 
$n_q+n_s$ independent and uniformly distributed
random angles $\phi_k$, with $1\leq k\leq n_q$, and $\phi'_k$,  
$1\leq k\leq n_s$, for some integer $n_s$.
Applying the operator 
$\prod_{k=n_s}^{1} \cnot_{i_{k}, j_{k}} 
\prod_{k=1}^{n_s}\left(R_{j_k}(\phi'_k) \cnot_{i_k, j_k}\right)
\prod_{k=1}^{n_q}R_k(\phi_k)$ multiplies each basis state
by a
random variable $\pm \phi_1\pm\phi_2\pm\cdots\pm\phi'_{n_s-1}\pm\phi'_{n_s}$,
which for large $n_s$ tends to a Gaussian random variable.
Here $R_j(\phi)=\exp(i\phi\sigma_j^{z}/2)$ 
is the rotation on the $j$-th qubit by an angle $\phi/2$,
$\cnot_{i,j}$ (controlled-not) is
the bit-flip on the $j$-th qubit conditioned by the $i$-th qubit, and
the $i_k$ and $j_k$ are chosen randomly between $0$ and $n_q-1$.
This transformation requires $(3n_s+n_q)$ gates instead of the $n_q^2$ 
gates needed for $ e^{-2i\pi\hat{p}^2/N}$. In practice, $n_s$ is taken
proportional to $n_q$, and the simulation requires in total
$n_q^2-n_q+2n_s$ two-qubit gates and $4n_q+n_s$ one-qubit gates.
This is quite close to what is needed to simulate the quantum 
baker's map already implemented \cite{cory}, thus
ISRM should be implementable as well
in present-day quantum computers. 

\begin{figure}
\begin{center} 
\vspace{0.5cm}
\includegraphics[width=.65\linewidth, angle=-90]{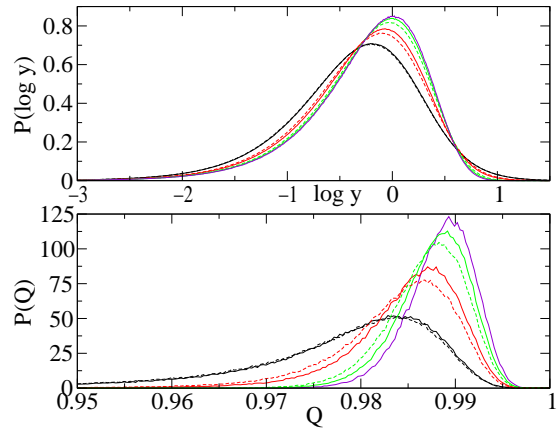} 
\end{center} 
\caption{(color online) Same as Fig.\ref{entangp2} for iterates
of ISRM, for $N=2^8$ (non-symmetric case). 
Averages are made over $1000$ disorder
realizations at $n=10^5$ (full lines), $n=1000$ (dashed lines, top)
and $n=2000$ (dashed lines, bottom). }
\label{entangphi}
\end{figure} 

\begin{figure}
\begin{center} 
\vspace{0.5cm}
\includegraphics[width=.85\linewidth]{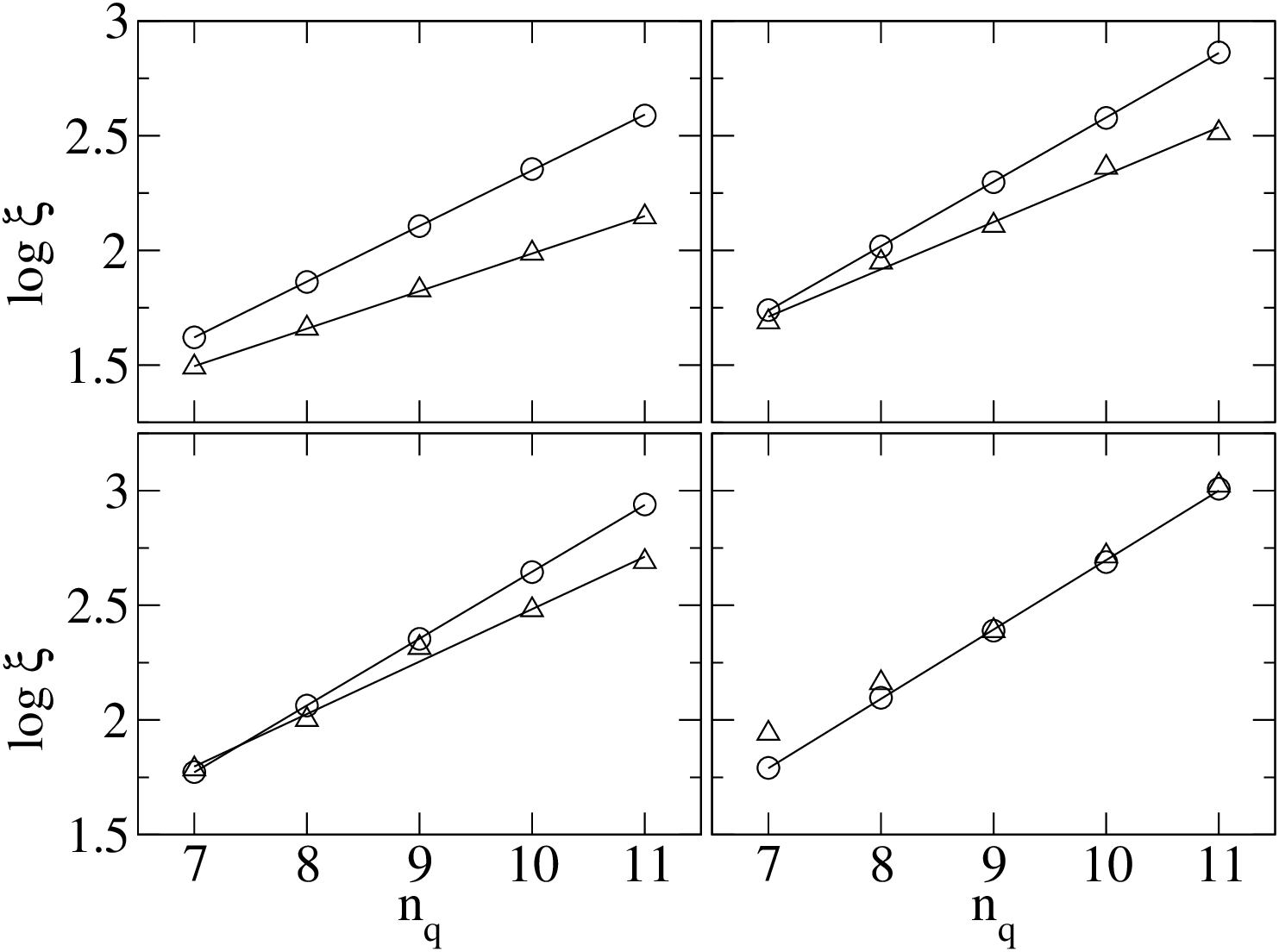} 
\end{center} 
\caption{IPR as a function of $N=2^{n_q}$ for column 
vectors (circles) and eigenvectors (triangles)
of ISRM (non-symmetric case), for $\alpha=1/3, 1/5$ (top),
$\alpha= 1/7, (1+\sqrt{5})/2$ (bottom). 
 Straight lines show $\xi\propto 
N^{0.80}$, $N^{0.93}$, $N^{0.97}$, 
$N^{1.00}$ (column vectors), and $\xi\propto N^{0.54}$, $N^{0.69}$, 
$N^{0.76}$, $N^{1.00}$ (eigenvectors). Logarithms are decimal.}
\label{ipr}
\end{figure} 

The latter algorithm 
can be used to generate pseudorandom operators having original properties.
Fig.\ref{entangphi}
shows that for ISRM matrix element and
$Q$ distributions converge
to a limiting distribution which depends on $\alpha$.  For
irrational $\alpha$, the distribution 
corresponds to CUE predictions (even in the symmetric case where
eigenvalue statistics follow COE, data not shown).
For rational $\alpha$ where eigenvalue statistics follow 
semi-Poisson predictions,
the limiting distributions are different 
from both COE and CUE distributions in $p$ representation (in $q$ 
representation they converge to CUE predictions, data not shown).
Intermediate eigenvalue statistics are also usually associated
with fractal properties of eigenvectors.
This was observed in the Anderson model at the metal-insulator
transition or in pseudointegrable systems. As was seen in 
\cite{bogomolny}, this is also
the case for the ISRM.
In a quantum information setting, randomly chosen 
eigenvectors of ISRM can be obtained by using the phase estimation
algorithm and measuring an eigenvalue: the wavefunction collapses
to the eigenvector associated to the eigenvalue measured.
One can also easily obtain column vectors of ISRM by iterating
a basis vector.
In Fig.\ref{ipr} we display the Inverse 
Participation Ratio (IPR) for the eigenvectors and
column vectors. This quantity is given by
$\xi = \sum_i|\Psi_i|^2/\sum_i|\Psi_i|^4$ for a wavefunction 
$|\Psi \rangle=\sum_i\Psi_i |i\rangle $.
It gives the number of basis states supporting the wavefunction  
($\xi=1$ for a state localized on a single basis state, and $\xi=N$ for 
a state uniformly spread over $N$ of them).  The results show that
where intermediate statistics are present, 
for both eigenvectors and column vectors of ISRM one has
$\xi \propto N^\gamma$
with $\gamma <1$, indicating fractal distributions of components. 

In conclusion, we have shown that 
quantum maps displaying intermediate statistics
can be simulated with a remarkable economy of resources on a quantum computer,
especially in a NMR setting.  We have also explored the link
between such intermediate statistics, entangling power of 
the quantum evolution and matrix elements distribution.  At last, we have shown
that a suitably randomized map can be used as an efficient generator
of pseudorandom operators displaying statistical properties different
from RMT, and in particular producing fractal random vectors.

We thank K. Frahm for discussions, 
and CalMiP in Toulouse and IDRIS in Orsay 
for access to their supercomputers.
This work was supported in part by the 
project EDIQIP of the IST-FET program of the EC.


\end{document}